# Search for Anomalously Heavy Isotopes of Helium in the Earth's Atmosphere

P. Mueller,[1] L.-B. Wang,[1,2] R. J. Holt,[1] Z.-T. Lu,[1] T. P. O'Connor,[1] J. P. Schiffer[1,3]

[1]*Physics Division, Argonne National Laboratory, Argonne, Illinois 60439*
[2]*Physics Department, University of Illinois at Urbana-Champaign, Urbana, Illinois 61801*
[3]*Physics Department, University of Chicago, Chicago, Illinois 60637*

Motivated by the theoretical hypotheses on the existence of heretofore unobserved stable elementary particles and exotic nuclear states, we searched for doubly-charged particles, as anomalously heavy isotopes of helium, in the Earth's atmosphere using a sensitive laser spectroscopy technique. The concentration of noble-gas-like atoms in the atmosphere and the subsequent very large depletion of the light $^{3,4}$He isotopes allow us to set stringent limits on the abundance: $10^{-13}$ - $10^{-17}$ per atom in the solar system, over the mass range of 20 – 10,000 amu.

Our knowledge of the stable particles that may exist in nature is defined by the limits set by measurements. As has been pointed out by Glashow *et al.* [1], this is an experimental result and there remain possibilities for 'superheavy' particles in the mass range of $10 - 10^5$ amu (atomic mass unit). The most interesting limits are those that can be set for singly charged [2] or neutral particles, but other charges are also of interest. There have also been suggestions that there may be very tightly bound stable states of hadronic matter, e.g. 'collapsed nuclei' [3] or 'abnormal states of nuclear matter' [4]. A more recent motivation for experimental searches is the possible existence of strange quark matter (so-called 'strangelets' with approximately equal numbers of up, down and strange quarks) that was first pointed out by Witten [5] and explored by Fahri and Jaffe [6], and more recently in [7][8][9]. The suggestion generally is that such states would have a lower charge-to-mass ratio than normal nuclei, and for the stranglets, very much lower (*e.g.* corresponding to a baryon number A somewhere in the region of between 20 and 80 amu for atomic number $Z = 2$, and around 50 to 200 amu for $Z = 8$).

A particularly favorable case is presented by particles of charge +2e, in other words helium-like particles. Normal helium is severely depleted in the terrestrial environment because of its light mass. The primordial helium finds its way to the exosphere, from where it escapes into space because of its low mass, and thus helium is replenished only from radioactive decay. Other noble gases, such as neon, argon, krypton, or xenon, are concentrated largely in the Earth's atmosphere, after an initial, lesser depletion relative to solar system levels at the early stages of the planet's evolution [10][11]. A heavy (mass > 20 amu) and doubly-charged particle would then be helium-like but behave as other noble gases and remain in the atmosphere. The concentration of noble-gas-like atoms in the atmosphere and the subsequent very large depletion of the known light $^{3,4}$He isotopes from the atmosphere allow significantly enhanced limits to be set.

Searches for tightly bound stable nuclei, such as strangelets, have been carried out with several techniques [12]. Some measurements have looked for possible pion emission when a thermal neutron is captured by such a system [13], a technique that is probably not applicable to strangelets. Mass spectrometry was used on a variety of light elements [14][15][16], and achieved the lowest limits (as low as ~ $10^{-30}$ on isotopic abundance at around 100 amu) on heavy isotopes of hydrogen with isotopically enriched heavy water [17]. This technique assumes that very heavy hydrogen would remain in water and not be segregated in solid precipitates, and would not be lost in the electrolysis enrichment process. Photon-burst spectroscopy has been used to search for Na-like strangelets and achieved an isotopic limit of $5 \times 10^{-12}$ in the range of $10^2 - 10^5$ amu [18]. The technique of searching for anomalous backscattering of heavy-ions by strangelets [19] and the technique of searching for anomalous high-energy γ rays emitted by strangelets upon heavy-ion activiation or (p, n) reactions [20] were used to set limits on the order of $10^{-10} – 10^{-17}$ in the mass range of $10^3 – 10^8$ amu. Searches specifically aimed at anomalous isotopes of helium have been performed in two previous experiments. It was presented in a conference that, with accelerator mass spectrometry, Klein *et al.* [15] set an isotopic abundance limit of $6 \times 10^{-15}$ over the mass range of 3.06 – 3.96 amu and 4.04 – 8.12 amu. With conventional mass spectrometry, Vandegriff *et al.* [21] extended the search to the mass range of 42 – 82 amu (Fig. 3).

In this work, we used a laser spectroscopy technique and took advantage of the isotope shift due to the higher mass of the heavy nucleus. The electronic structure of such an abnormal helium atom should be identical to that of ordinary helium; the influence of the nucleus is reflected only in small isotope shifts and in hyperfine structure. In helium as in all light elements the isotope shift is dominated by the so-called mass shift that results from the change of the nuclear mass. The change in the charge distribution of the nucleus that leads to the so-called field shift is many orders of magnitude smaller and can be neglected here. The mass shift $\delta\nu_{MS}$ of a transition between isotopes of nuclear mass *M* and mass infinity is given as

$$\delta\nu_{MS} = -\frac{F_{MS}}{M}.$$

The mass shift constant $F_{MS}$ can be extracted either from experimental isotope shift measurements or from theoretical calculations. This technique is particularly suited to searches for isotopes with unknown mass because the range of atomic transition frequency to be searched is finite even as the atomic mass goes to infinity. We performed the search by probing the 1s2s $^3S_1 \rightarrow$ 1s2p $^3P_2$ transition at 1083 nm in helium atoms at the metastable 1s2s $^3S_1$ level. From the known $^3$He - $^4$He isotope shift [22], it is derived that $F_{MS}$ = 412 GHz for this transition. It is noted that this search method can be applied not only to anomalously heavy isotopes of helium but also to such searches in other atomic or molecular species.

The helium sample was extracted from air with sorption pumps cooled to 80 K by liquid nitrogen, thus effectively absorbing all major gases in air except neon, helium and hydrogen, whose partial pressures in air are 13.8 mTorr, 4.0 mTorr and 0.4 mTorr, respectively [23]. After the sorption pumps have reached equilibrium the remaining gas was compressed with a turbopump into a previously evacuated quartz glass cell for laser spectroscopy work. In the cell the chemically active gases such as hydrogen and water were absorbed by a getter pump, leaving only neon and helium with the pressure ratio of approximately 3.5:1, as measured with a residual gas analyzer and consistent with the well-known atmospheric value. The gas purity of the sample proved to be critical, as any impurities except neon can quench the metastable state of helium upon collisions.

The process of freezing out the gases is governed by the surface adsorption time,

$$\tau = \tau_0 e^{E_s/k_B T},$$

where $E_s$ is the binding energy, $\tau_0$ is the period of oscillation of helium atoms temporarily bound to the surface, $k_B$ is the Boltzmann constant, and $T$ is the temperature of the surface. $E_s$ is determined mainly by the electronic structure which depends weakly on nuclear mass. Existing measurements of helium atoms on various surfaces indicate that $E_s/k_B < 20$ K [24]. Therefore, at the sorption-pump temperature of 80 K, $e^{E_s/k_B T} \sim 1$ even for the heaviest anticipated masses. The oscillation period $\tau_0$ is proportional to $\sqrt{m}$, where $m$ is the mass of the atoms. On the other hand, the rate of atoms hitting the surface is proportional to the velocity of the atoms in the vapor, and is proportional to $1/\sqrt{m}$. These two factors cancel and, consequently, the ratio of helium atoms in the vapor to that on the surface should only be weakly dependent on mass. Thus, we conclude that the heavy helium was extracted with approximately the same efficiency as the ordinary helium. A practical high-mass limit in the search, set at $1 \times 10^4$ amu, is introduced by the uncertainty in spectroscopic lineshape which will be discussed below. At 80 K such heavy helium atoms have a $1/e$ height of 5 m, much larger than the typical height (0.5 m) of the apparatus. We also note that, within the above mass range, convection keeps the vertical distribution of masses in the atmosphere near ground reasonably uniform.

A schematic diagram of the apparatus is shown in Fig. 1. In order to search for a weak absorption signal and avoid Doppler broadening, we performed frequency-modulation saturation spectroscopy [25] on the 1s2s $^3S_1 \rightarrow$ 1s2p $^3P_2$ transition (natural linewidth = 1.6 MHz). The gas sample was enclosed in a 1 m long and 2.5 cm diameter quartz glass cell, in which a RF-driven discharge was used to populate the metastable 1s2s $^3S_1$ level *via* electron-impact collisions. The gas pressure in the cell required a judicious choice between optimum metastable population and higher sample concentration. Tests indicated that an optimum pressure occurs at approximately 200 mTorr, which provided an estimated metastable population of around $1 \times 10^{-4}$ relative to groundstate helium atoms. A laser system consisting of two diode-lasers and a fiber amplifier provided the required laser power of 500 mW and the single-mode laser frequency with long-term stability and scan control of better than 1 MHz. The probe laser beam was phase-modulated at 36 MHz with an electro-optical modulator, directed through the long cell with a power of 8 mW and a diameter of 1 cm, and finally focused onto a fast InGaAs-photodiode detector. The pump beam was frequency shifted by 2 MHz and chopped at 45 kHz with two acousto-optical modulators, then directed through the long cell in the opposite direction with a power of 50 mW and a diameter of 1 cm. This 2 MHz shift was implemented to avoid interference effects between the probe and pump laser beams, yet was small enough so that the frequencies of the pump and probe beam still overlapped within the Doppler width of the heavy helium atoms. The signal from the photodiode was first demodulated at 36 MHz with a RF frequency mixer and then at 45 kHz with a lock-in amplifier, whose output was recorded as data.

The calibration of the detection sensitivity was accomplished by using $^3$He as a reference isotope, whose abundance in air is $1.4 \times 10^{-6}$ relative to $^4$He. We have also calibrated our method against mass spectrometry using three different samples whose $^3$He isotopic abundance ranged from $10^{-7}$ to $10^{-5}$ [26]. At the same strong-discharge condition used for the search for a heavy isotope, the observed signal of $^3$He in the atmospheric sample had a signal-to-noise ratio of 70 and a linewidth of 45 MHz, which was significantly broadened by the high pressure and high laser power. The noise was mainly due to laser power fluctuations and was approximately a factor of 5 larger than the shot noise. The $^3$He lineshape was important as it was used as a template when looking for a signal and it was recorded every day during the search in order to verify that all system parameters were set correctly. For the search the laser was slowly scanned in the range of 20 - 109 GHz above the transition frequency of $^4$He, which corresponds to a mass range of 5 amu – infinity. The laser-frequency control system allowed for a continuous scan of a maximum of 320 MHz. The complete search was therefore split into many individual scans, each with a width of 320 MHz, that overlapped by 20 MHz on each side.

A first analysis of the data showed that the observed fluctuation in the signal was not statistical noise but a spurious signal caused by an etalon effect in the probe beam with a free spectral range of approximately 400 MHz, much larger than the 45 MHz linewidth of the sought-after signal. The spurious sinusoidal signal was filtered out with only a 3% reduction in detection sensitivity, as demonstrated by a comparison of the $^3$He signal before and after the filtering. Moreover, based on the $^3$He signal, it was verified that the first derivative of a Lorentzian peak with a width of 45 MHz was an appropriate template (Fig. 2b). The filtered data (Fig. 2a) were used to search for a possible signal by sliding the template over the scan range and calculating the best fit for the amplitude for each frequency with a resolution of 3 MHz (Fig. 2c). We find the distribution of the fit amplitudes over the entire frequency range to be statistical with the mean value of zero (Fig. 2d). We conclude at the 95% confidence level that there is no anomalous peak with an amplitude larger than $7.9 \times 10^{-2}$ times the $^3$He amplitude anywhere in the entire frequency range.

This detection method is more sensitive to atoms of heavier masses because the Doppler width decreases and the fraction of the atoms being resonantly probed increases with the atomic mass until the Doppler width becomes comparable to the interaction linewidth of 45 MHz as the mass reaches $1 \times 10^4$ amu. For masses above this value, the shrinking Doppler width may in fact alter the signal linewidth and reduce the detection sensitivity. For this reason we conservatively set the high-mass limit at $1 \times 10^4$ amu. The sensitivity depends slightly on the hyperfine structure due to the difference in population distribution and transition strength. For example, the signal of $^3$He via the $2^3S_1$ F=3/2 → $2^3P_2$ F=5/2 transition is a factor of 1.7 weaker than that of the corresponding transition in $^4$He. The quoted limits assume that the heavy helium atom has no hyperfine structure with the understanding that the limits will change by a small factor if hyperfine structure is present. The limits on the isotopic abundance of heavy doubly-charged particles in the Earth's atmosphere are shown in Fig. 3.

The origin and evolution of the terrestrial noble gases have been subjects of great interest in geology for the past five decades [11][27]. Although it is not conclusively established, past geochemical measurements indicate, and most global evolution models infer or assume, that the total inventory of primordial noble gases in the mantle is small compared to that in the atmosphere. For the case of helium, it is known that the $^{3,4}$He nuclei present in the Earth are almost all "young" nuclei with radiogenic or cosmogenic origin, and that the primordial $^{3,4}$He escaped into space because the lifetime for helium in the atmosphere (the result of diffusion from the mantle) is only ~ $2 \times 10^6$ years. Knowing that there are $6 \times 10^{38}$ $^4$He nuclei in the Earth's atmosphere and a total of $1 \times 10^{50}$ nuclei of all kinds in the Earth [28] (composition in terms of atomic abundance: O, 50%; Fe, 18%; Mg, 17%; Si, 14%), we can set limits on the abundance of doubly charged particles in the whole Earth (Fig. 3).

It is believed that the sun and the planets formed from the same starting material, and that this original composition is preserved in the sun. The noble gases, as well as hydrogen, were either not captured in the planet formation process, or were subsequently depleted in the Earth at the early stage when the planet was molten. The deficiency factors for each noble gas element (Fig. 4), defined as the ratios of the abundances of the elements in the Earth over those in the Sun, are well documented [27][29]. There is clearly a mass dependence: heavier noble gas atoms are retained more than the lighter ones, and for atoms of mass > 80 the deficiency factor approaches a constant. Assuming that the deficiency factors for the anomalous helium follow the same mass dependence, we can set limits on their abundance in the solar system (Fig. 3).

In conclusion, we have employed a sensitive laser spectroscopy method to search for doubly-charged particles with anomalously high mass in the Earth's atmosphere. Compared with the previous searches of such particles with mass spectrometry, this work significantly extends our knowledge over a wider range of mass and with much improved limits. The sensitivity of this method could be improved further by perhaps several orders of magnitude with the application of cavity-enhanced spectroscopy [30] and by perhaps enriching heavy helium with gas chromatography.

We thank R. N. Boyd, J. Specht and N. Sturchio for stimulating and helpful discussions, Y. Sano for providing $^3$He samples calibrated with mass spectrometry, K. Bailey for general technical support, and J. Gregar for making the long glass cell. This work is supported by the U.S. Department of Energy, Nuclear Physics Division, under contract W-31-109-ENG-38.

**Figures:**

1. Schematic diagram of the apparatus used for FM saturation spectroscopy. AOM, acousto-optical modulator; DL, diode laser; EOM, electro-optical modulator; FA, fiber amplifier; PBS, polarization beam splitter; PD, photo diode detector; RF, radio-frequency generator; FG, function generator.
2. (a) FM spectroscopy data. The frequency range of –85.8 - 0 GHz corresponds to the mass range of 4.8 – 10,000 amu. The region from –77.1 GHz to –69.7 GHz, corresponding to 5.3 – 5.9 amu, was omitted due to the interference of the $2^3S_1 \rightarrow 2^3P_0$ transition of $^4$He. (b) FM spectroscopy data of $^3$He. Note that both the vertical and the horizontal scales change from Fig. 2(a) to 2(b). The stronger peak is due to the $2^3S_{1, F=3/2} \rightarrow 2^3P_{2, F=5/2}$ transition and the weaker peak, 221 MHz away, is due to the $2^3S_{1, F=1/2} \rightarrow 2^3P_{2, F=3/2}$ transition. (c) Relative amplitudes obtained by fitting the data with the $^3$He template. The amplitudes are normalized to the amplitude of $^3$He whose isotopic abundance is $1.4 \times 10^{-6}$. Only 5% of the amplitudes lie beyond each dashed line. (d) Distribution of the amplitudes. Mean = $4.4 \times 10^{-5}$; standard deviation = $1.4 \times 10^{-2}$; maximum amplitude (= $5.6 \times 10^{-2}$) occurred at the laser frequency of –27.0 GHz (or 15.3 amu); 95% of the amplitudes are above $-2.3 \times 10^{-2}$. The 95% confidence limit is set to be $7.9 \times 10^{-2}$ (= $5.6 \times 10^{-2} + 2.3 \times 10^{-2}$).
3. Limits on the abundance of anomalously heavy isotopes of helium. The solid lines indicate limits set by this work: top, on the isotopic abundance (anomalous helium vs. $^4$He) in the Earth's atmosphere; middle, on the atomic abundance (anomalous helium vs. total number of nuclei) in the solar system; bottom, on the atomic abundance in the Earth. The dashed line indicates the limits set by Vandegriff *et al*. [21] on the atomic abundance in the solar system.
4. Deficiency factors of terrestrial noble gas isotopes [27]. The deficiency factor of an isotope is defined as the ratio of the atomic abundance of a certain element in the earth over that in the sun. The terrestrial $^4$He has a radiogenic origin. A fit (dotted line) over the four data points of primordial isotopes yield that $\log(Def.) = -3.1 - 9.0 \cdot \exp(-0.045 \cdot M)$. This function is used to calculate the solar abundance from the terrestrial abundance of heavy helium.

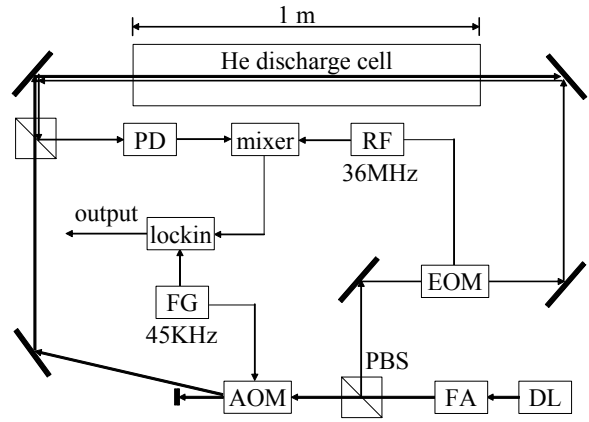

Figure 1

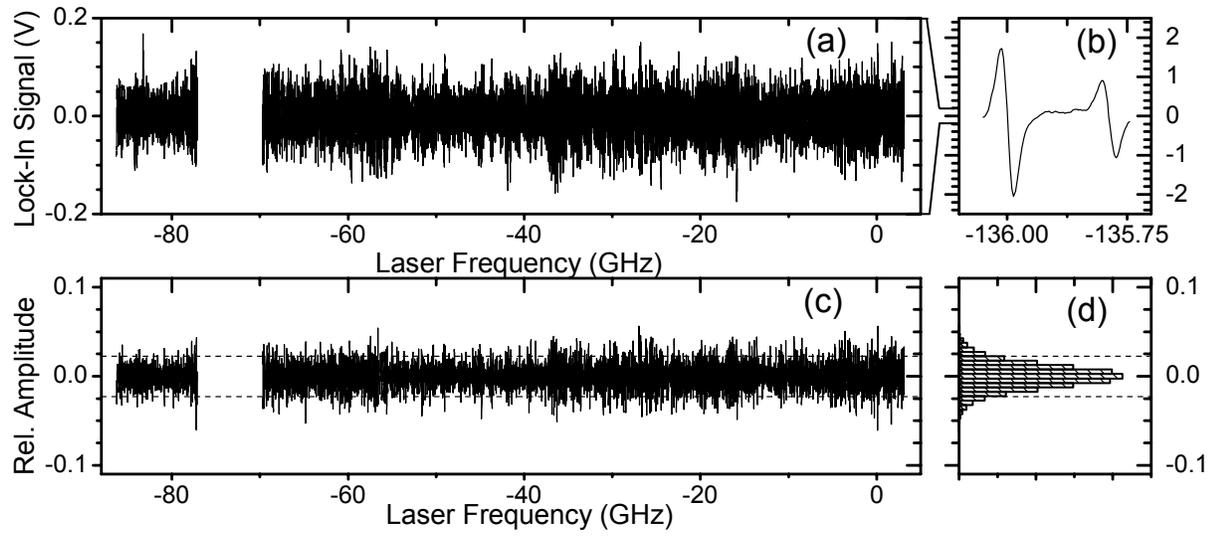

Figure 2

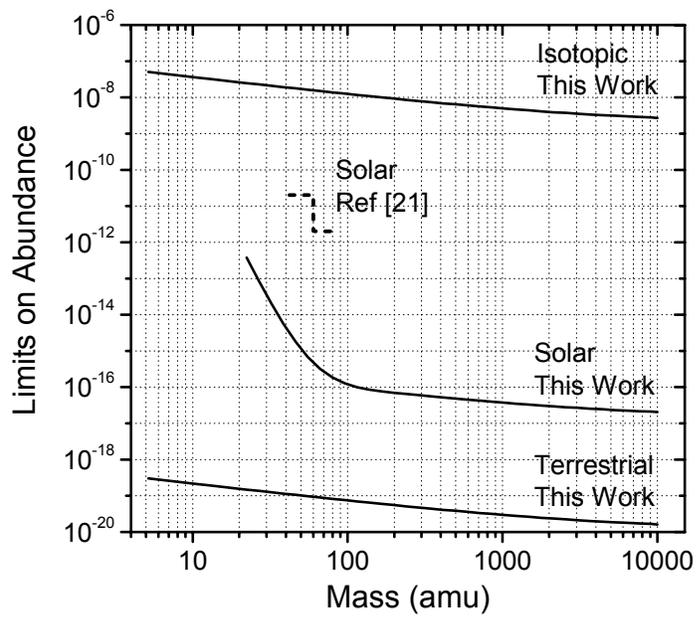

Figure 3

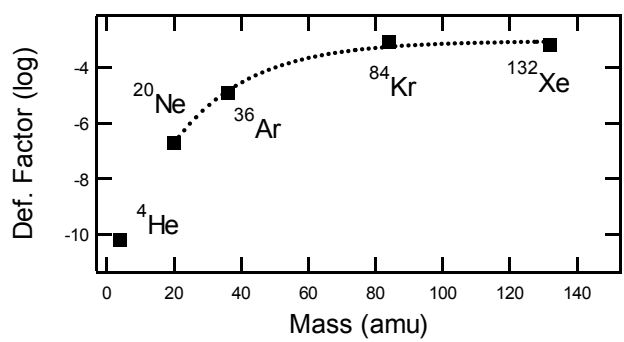

Figure 4